\documentstyle[epsf,12pt]{article}
\begin{document}
\begin{center}
\Large{\bf A Heavy Ion Fireball freeze-out Dipion Cocktail for Au-Au Collisions at $\sqrt{s_{NN}}$=200 GeV (Part 1).}\\
\large{R.S. Longacre$^a$\\
$^a$Brookhaven National Laboratory, Upton, NY 11973, USA}
\end{center}
 
\begin{abstract}
In this paper we develop all the ingredients that come into play in the 
freeze-out of the heavy ion fireball for Au-Au collisions at 
$\sqrt{s_{NN}}$=200 GeV into dipions. The resonance production of particles 
that decay into dipions plus minijets that also decay into dipions are 
explored. The final state re-scattering of the pions from minijets play an 
important role in the mass spectrum of the cocktail. Mass shifts due to minijet
interference which depend on the volume of the re-scattering pions is 
presented. The effective mass balance function is explained. The dipion mass 
spectrum within a $p_t$ range(intermediate $p_t$) is fitted using thermal and
minijet amplitudes.
\end{abstract}
 
\section{Introduction} 

The ultra-relativistic heavy ion collision starts out as a state of high 
density nuclear matter called the Quark Gluon Plasma(QGP) and expands rapidly 
to freeze-out. During the freeze-out phase quarks and gluons form a system of 
strongly interacting hadrons. These hadrons continue to expand in a thermal 
manner until no further scattering is possible because the system becomes to 
dilute. However this transition from quarks and gluons(partons) into hadrons is
not a smooth affair. The expansion is very rapid and some faster or hard 
scattered partons fragment directly into hadron through a minijet\cite{Trainor}
process. Thus we have thermal and minijet hadrons present in the last 
scattering of the hadrons. This mixture of sources is considered in this paper 
and applied to the dipion mass spectrum of the heavy ion fireball formed in 
Au-Au collisions at $\sqrt{s_{NN}}$=200. 

The paper is organized in the following manner:

Sec. 1 is the introduction to the cross sections for $\pi$ $\pi$ scattering. 
Sec. 2 develops a two component model with direct production and decay plus 
a background $\pi$ $\pi$ component which must re-scatter in order
to satisfy unitarity. Sec. 3 consider two channel unitary scattering 
$\pi$ $\pi$ and $K$ $\overline{K}$ for $J^{PC}$ = $0^{++}$ and shows what 
re-scattering would look like in the $\pi$ $\pi$ channel. Sec. 4 introduces 
the balance function for dipion data and applies it to the two component model.
Sec. 5 applies the two component model to dipion data within a $p_t$ range.
Sec. 6 presents the summary and discussion. Finally there are two appendices. 
Appendix A show the mathematical details of how the two component model is 
worked out. Appendix B determines the maximum value of the $\alpha$ parameter 
using photo-production data reported in 
Ref.\cite{photo_rho}. 

\subsection{$\pi$ $\pi$ scattering cross section} 

For the first part of this story we will define what a scattering
cross section is. We will first only consider elastic
scattering of pions. Two pions can scatter at a certain energy
which we will call $M_{\pi \pi}$. The differential cross section
$\sigma$ at a given $M_{\pi \pi}$ is
\begin{equation}
\frac{d\sigma}{d\phi d\theta} = \frac{1}{K^2} \left| \sum_{\ell}
(2\ell + 1) T_{\ell} P_{\ell} (cos\theta) \right| ^2
\end{equation}
where $\phi$ and $\theta$ are the azimuthal and scattering angles,
respectively. $T_{\ell}$ is a complex scattering amplitude and
$\ell$ is the angular momentum. $P_{\ell}$ is the Legendre
polynomial, which is a function of $cos\theta$. $K$ is the flux
factor equal to the pion momentum in the center of mass. The $T_{\ell}$
elastic scattering amplitudes are complex amplitudes
described by one real number which is in units of angles. The form
of the amplitude is
\begin{equation}
T_{\ell} = e^{i \delta_{\ell}} sin\delta_{\ell}
\end{equation}
We note that $\delta$ depends on the value of $\ell$ and $M_{\pi
\pi}$. We will use the phase shifts given in Ref\cite{Grayner}.

\section{The two component model}
Let us consider two pions scattering in the final state of the
heavy ion collision. The scattering will be in some $\ell$ 
partial wave. The $M_{\pi \pi}$ of the scattering
dipion system will depend on the probability of the phase space
of the overlapping pions. The pions emerge from a close encounter
in a defined quantum state with a random phase. We will call this
amplitude $A$ and note that the absolute value squared of the
amplitude is proportional to the phase space overlap. The emerging
pions can re-scatter through the quantum state of the pions, which is 
a partial wave or a phase shift. We have amplitude $A$ plus $A$ times 
the re-scattering of pions through the phase shift consistent quantum state of 
$A$. The correct unitary way to describe this process is given by 
Ref\cite{Aaron} equation(4.5)
\begin{equation}
T = \frac{V_1 U_1}{D_1} + \frac{\left( V_2 + \frac{D_{12}
V_1}{D_1} \right)\left( U_2 + \frac{D_{12} U_1}{D_1}\right)}{D_2 -
\frac{D_{12}^2}{D_1}}
\end{equation}
In the above equation we have two terms, 1 and 2. The first term
denoted by 1 is the $\pi \pi$ scattering through $\ell$-wave which will
become the amplitude $A$ mentioned above, where $V$ is the incoming
and $U$ is the outgoing $\pi \pi$ system. The second term denoted
by 2 is the direct production of the $\pi \pi$ system in the $\ell$-wave
with $V$ being the production, the propagation being $D$ and the decay 
being $U$. We see that there are terms $D_{12}$ which involves a loop of pions
between scattering pions and the formation of a resonance($\ell$ =1 would be a
$\rho$) by the pions.

\subsection{Final equation for the two component model}
The complete derivation is in Appendix A. From the appendix we
get two terms, one being the direct production of the resonance or
$\pi \pi$ $\ell$-wave phase shift and the second being the resonance from
re-scattering. The final equation 6 has two important factors,
one is two-body phase space and the other is a coefficient $\alpha$.
This coefficient is related to the real part of the $\pi \pi$ re-scattering
loop and is given by equation 4. When the pions re-scatter or interact at
a close distance or a point the real part of $\alpha$ has its maximum value of 
$\alpha_0$. While if the pions re-scatter or interact at a distance determined
by the diffractive limit the value of $\alpha$ is zero.

In the equation 6 $|T|^2$ is the cross section for $\ell$ partial wave
produced, where $D$ is the direct production amplitude and $A$
is the amplitude introduced above for the re-scattering pions into
the $\ell$-wave with $\delta_{\ell}$ the $\pi \pi$ phase shift\cite{Grayner}.
The $q$ is the $\pi \pi$ center of mass momentum. At a given $p_t$ and $y$ bin,
$D$ will have a thermal factor as a function of $M_{\pi \pi}$. The $\alpha$
which is the real part of the re-scattering factor has a simple form given by
\begin{equation}
\alpha = (1.0 - \frac{r^2}{r_0^2}) \alpha_0
\end{equation}
where $r$ is the radius of re-scattering in fm's and $r_0$ is 1.0
fm or the limiting range of the strong interaction ranging to  $r$ = 0.0 for 
point like interactions.

The dependence of $A$ is calculated by the phase space overlap of dipions 
added as four vectors and corrected for proper time, with the sum
having the correct $p_t$ , $y$ and phase space weighting for $\ell^{th}$ 
partial wave for a given $M_{\pi \pi}$.

Finally we must use the correct two body phase space. For a two body
system of pions, phase space goes to an constant as $M_{\pi \pi}$ goes
to infinity. Let us choose this constant to be unity. Phase space which
is denoted by PS is equal to
\begin{equation}
PS = \frac{2qB_{\ell}(q/q_s)}{M_{\pi \pi}}
\end{equation}
where $B_{\ell}$ is a Blatt-Weisskopf-barrier factor\cite{Hippel} for $\ell$
angular momentum quantum number. The $q_s$ is the momentum related to the
range of interaction of the $\pi \pi$ scattering. 1 fm is the usual
interaction distance which implies that $q_s$ is .200 GeV/c. For the $\rho$
meson $\ell$=1 the barrier factor is $B_1$ = 
$\frac{(q/q_s)^2}{(1+(q/q_s)^2)}$. The phase space factor PS as a function of 
$q$ near the $\pi \pi$ threshold is given by $q^{2\ell + 1}$. Thus in the 
appendix we use $q^{2\ell + 1}$ for the factor PS except for equation 6 which 
is the final equation.

\begin{equation}
|T|^2  = |D|^2 \frac{sin^2\delta_{\ell}}{PS} +
\frac{|A|^2}{PS} \left| \alpha sin\delta_\ell + PS cos\delta_\ell \right|^2
\end{equation}

\section{Re-scattering through the Swave $\pi \pi$ is a two channel problem}

In Sec. 2 we derived equation 6 considering only elastic scattering of the
$\pi \pi$ system. If we consider the Dwave it couples to the $f_2(1270)$ 
($J^{PC}$ = $2^{++}$) with 85\% of the cross section in the $\pi \pi$ channel. 
The Pwave couples to the $\rho(770)$ ($J^{PC}$ = $1^{--}$) where 100\% is in 
the $\pi \pi$ channel. The Swave $\pi \pi$ ($J^{PC}$ = $0^{++}$) couples to 
two resonances the $\sigma$ and the $f_0(980)$. The $\sigma$ is purely elastic 
while the $f_0(980)$ is split between the $\pi$ $\pi$ and $K$ $\overline{K}$ 
channels. These two channels plus two resonances gives an additional complexity
to the re-scattering problem.

In order to handle the $\pi \pi$ and $K$ $\overline{K}$ channels we will use
the K-matrix approach. When we are below the $K$ $\overline{K}$ threshold
the system is only a one channel problem and the K-matrix is only a single
term of the matrix

\begin{equation}
K_{11} = tan\delta_0.
\end{equation}

We see that when $\delta_0$ = $90^\circ$ that there will be a pole in the
K-matrix. It is standard to expand the K-matrix as a sum of poles.

\begin{equation}
K_{11} = \sum_i {\frac{2\gamma_i^2q_{\pi \pi}}{M_{\pi \pi}}\over{(M_i^2 - M_{\pi \pi}^2)}} 
\end{equation}

Where $\gamma_i$ is the coupling of the pole($i^{th}$) to the $\pi \pi$ 
channel, $q_{\pi \pi}$ is the center of mass momentum of the $\pi \pi$ 
channel and $M_i$ is the mass of the pole($i^{th}$). The T-matrix is given by

\begin{equation}
T_{11} = e^{i \delta_0} sin\delta_0 = {K_{11}\over{(1 - iK_{11})}} = \,(1 - iK)^{-1}K
\end{equation}

When both channels are open $j=1$ $\pi$ $\pi$ and $j=2$ $K$ $\overline{K}$,
the K-matrix is given by

\begin{equation}
K_{11} = \sum_{i} {\frac{2\gamma_{i1}^2q_1}{M_1}\over{(M_i^2 - M_1^2)}},
K_{21} = \sum_{i} {\frac{2\gamma_{i2}\sqrt{q_2}\gamma_{i1}\sqrt{q_1}}{M_1}\over{(M_i^2 - M_1^2)}},
K_{12} = \sum_{i} {\frac{2\gamma_{i1}\sqrt{q_1}\gamma_{i2}\sqrt{q_2}}{M_1}\over{(M_i^2 - M_1^2)}},
K_{22} = \sum_{i} {\frac{2\gamma_{i2}^2q_2}{M_1}\over{(M_i^2 - M_1^2)}}.
\end{equation}

The T-matrix is given by

\begin{equation}
T  = \,(\delta - iK)^{-1}K.
\end{equation}

We fit the Swave ($J^{PC}$ = $0^{++}$) $\pi \pi$ of Ref\cite{Grayner} using
three poles for the $\sigma$, $f_0$ and some background from higher mass poles.
The $T_{11}$ amplitude is shown in Figure 1.

\begin{figure}
\begin{center}
\mbox{
   \epsfysize 6.8in
   \epsfbox{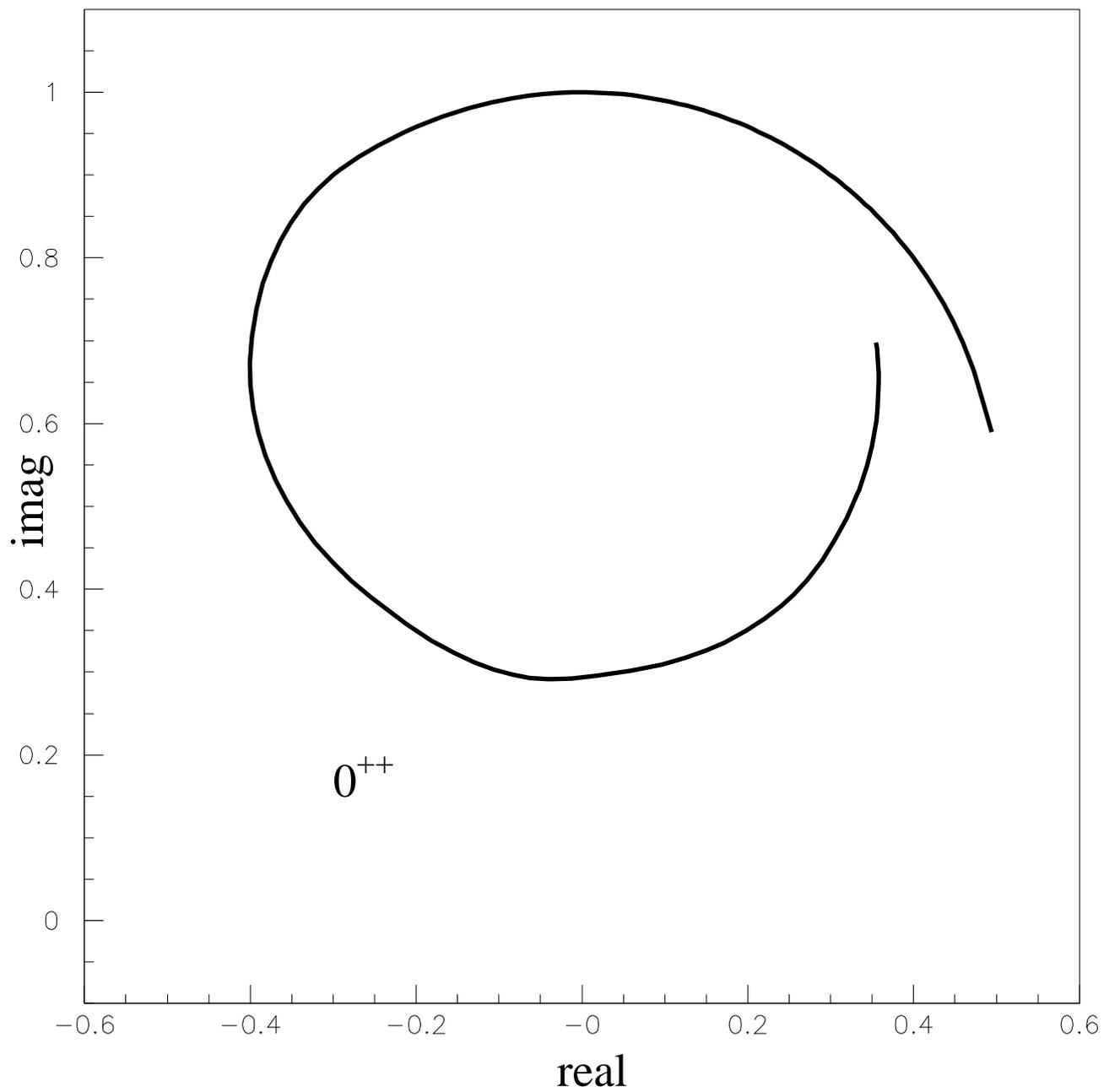}}
\end{center}
\vspace{2pt}
\caption{The $T_{11}$ amplitude for the Swave ($J^{PC}$ = $0^{++}$) 
$\pi \pi$ comes from a fit to of Ref\cite{Grayner} using three K-matrix 
poles for the $\sigma$, $f_0$ and some background from higher mass poles.}
\label{fig1}
\end{figure}

\clearpage

Two pions scattering in the final state of the heavy ion collision in a Swave
will be our amplitude $A$, where the emerging pions can re-scatter through the 
Swave. We have amplitude $A$ plus $A$ times the re-scattering of pions through 
the Swave phase shift. 

The term of equation 6 $PS e^{i \delta_0} cos\delta_0$ is equal to 
$PS(1 + ie^{i \delta_0} sin\delta_0)$. $T_{11}$ which is equal
to $e^{i \delta_0} sin\delta_0$ for the one channel case becomes  $T_{11}$ 
= $\eta e^{i \delta_0} sin\delta_0$ for the two channel case. Thus the 
re-scattering term becomes $PS(1 + i \eta e^{i \delta_0} sin\delta_0)$ or 
$PS(1 + iT_{11})$. Using our K-matrix fit to the Swave one obtain the 
re-scattering term plotted in Figure 2. In the lower mass we see a shift of the
spectrum to a lower mass. In the next section will see the same effect for
$\rho(770)$ resonance that re-scattering will shift its mass to lower values.
This shift will be caused by a direct production plus the re-scattering adding
together creating a shifted $\rho(770)$\cite{Fachini}. We see that the
$f_0$ is a narrow resonance. The $f_0$ resonates at the $K$ $\overline{K}$ 
threshold. Direct production of the $f_0$ gives a bump at the $K$ 
$\overline{K}$ threshold and the re-scattering of $\pi \pi$ also gives such a 
bump at the $K$ $\overline{K}$ threshold(Figure 2). Therefore we will only 
consider the $f_0$ as a resonance being directly produced and decaying into 
$\pi \pi$ near the $K$ $\overline{K}$ threshold.

\section{The balance function for dipion effective mass}

Up to this point in the paper we did not specify the charge of the pions
considered. With the idea of the balance function we look at the creation of
pairs of opposite charge pions. The QGP fireball begins mostly neutral without
a large excess of charge. Pairs of quarks and anti-quarks are created finally
forming hadrons mainly pions. Thus for every $\pi^+$ there is a $\pi^-$
which balance out the charge. The balance function measures the kinematic
variable between the balancing charge\cite{Pratt}. In our case we want to 
measure effective mass of the $\pi^+ \pi^-$ pairs. The dipion effective mass
balance function is defined by
\begin{equation}
B(M_{\pi\pi})  = \frac{1}{N}\sum_{events} \sum_{pairs} \frac{1}{2} \biggl\lbrace \frac{(M_{\pi^+\pi^-} - M_{\pi^+\pi^+})}{N_+} + \frac{(M_{\pi^+\pi^-} - M_{\pi^-\pi^-})}{N_-} \biggr\rbrace ,
\end{equation}
where $N$ is the number of events, $N_+$ is the number of positive pions per 
event, and $N_-$ is the number of negative pions per event. 

Let us consider the effective mass dipion balance function for p-p collisions
$\sqrt{s_{NN}}$=200 GeV\cite{Westfall}. The effective mass shows a large bump
at the $K^0$ mass from $K^0_S \rightarrow \pi^+ \pi^-$. Above this bump  there 
is another mass bump which could be the $\rho(770)$. At this mass Pwave 
should be the most important partial wave so let us assume this is 
the case. The balance function drops off with mass so we assume an
exponential fall off with $M_{\pi^+ \pi^-}$. The $\rho(770)$ should be 
directly produced and decay through the Pwave phase shift. We will use 
equation 6 to fit the p-p balance function. Figure 3 shows the fit to the 
STAR data, where the solid line is the fit and dashed line is the $\rho(770)$.
The exponential is amplitude A and the dotted line is $A$ plus $A$ times 
the re-scattering of pions through the Pwave phase shift. The value of $\alpha$
is 0.37 which means the radius of re-scattering of the pions is 0.9 fm using
equation 4 with the value of $\alpha_0$ equal to 2. The solid line which is a 
good fit to the data peaks at .73 GeV/c. We see that the shift of this peak is 
caused by the dotted line added to the dashed line. 

\begin{figure}
\begin{center}
\mbox{
   \epsfysize 6.8in
   \epsfbox{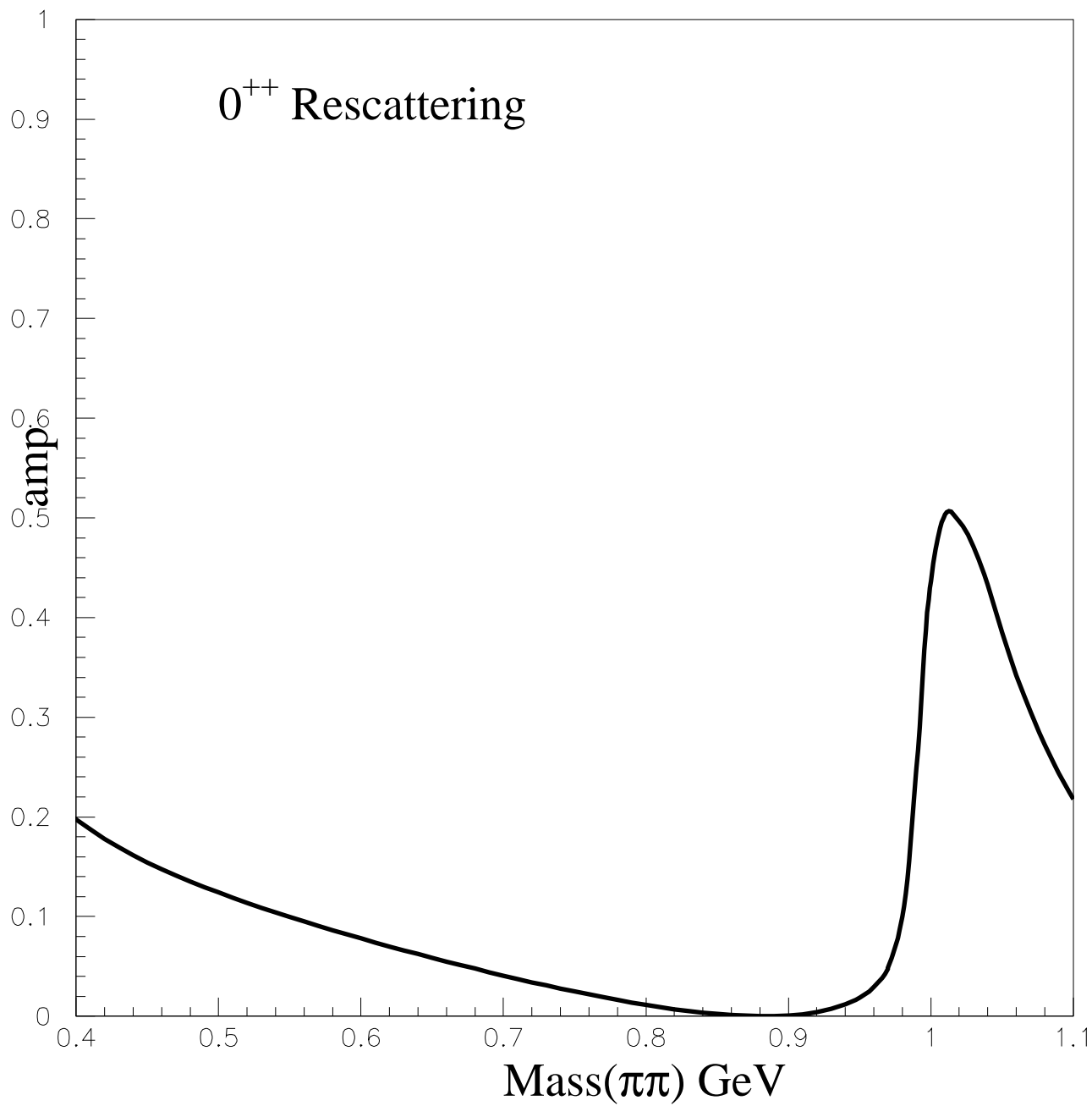}}
\end{center}
\vspace{2pt}
\caption{We plot for the Swave ($J^{PC}$ = $0^{++}$) $\pi \pi$ re-scattering
term $\left|PS(1 + iT_{11})\right|^2$. The $T_{11}$ amplitude comes from a fit 
to of Ref\cite{Grayner} using three K-matrix poles for the $\sigma$, $f_0$ 
and some background from higher mass poles.}
\label{fig2}
\end{figure}
 
\begin{figure}
\begin{center}
\mbox{
   \epsfysize 6.8in
   \epsfbox{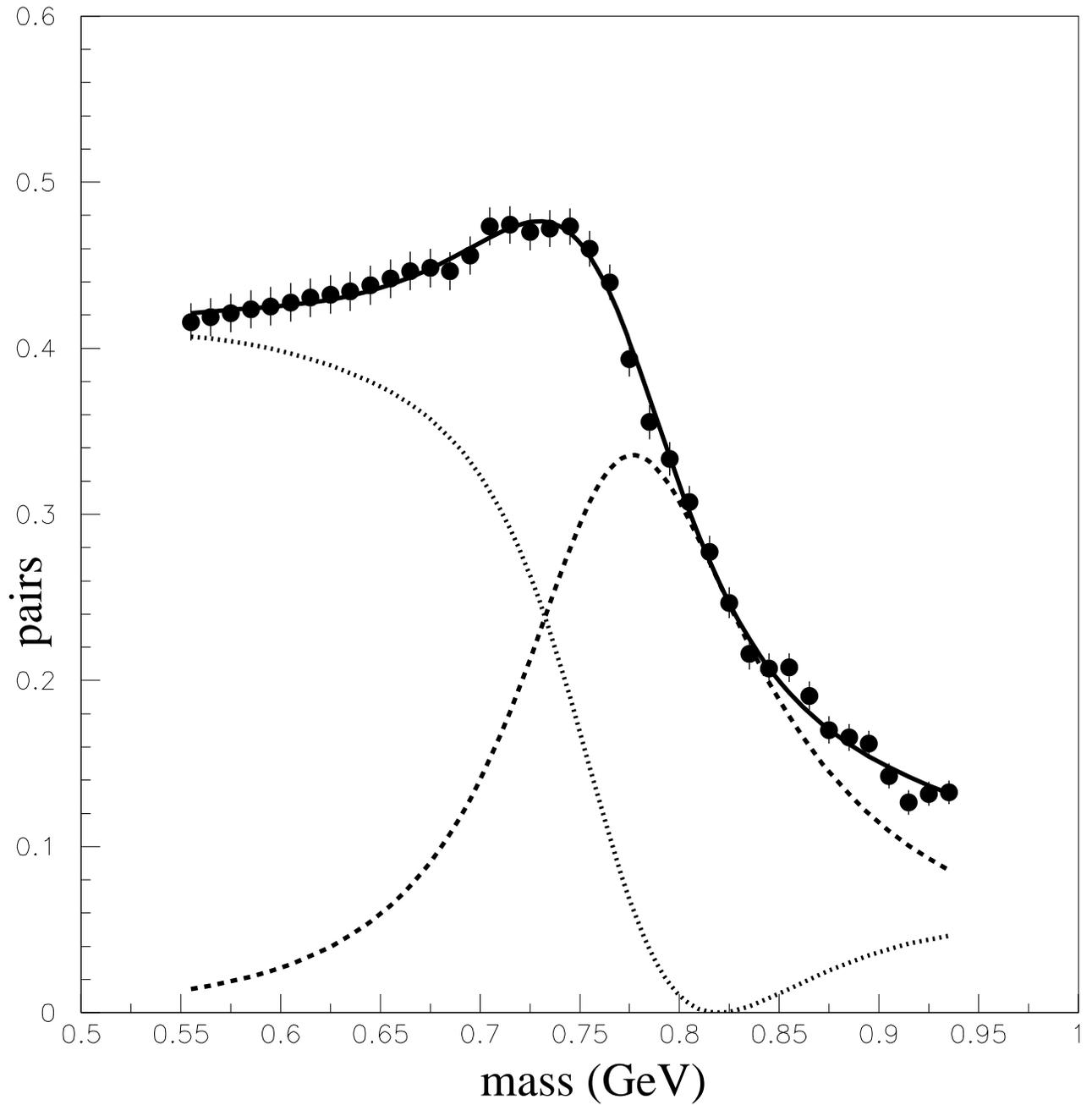}}
\end{center}
\vspace{2pt}
\caption{The fit to the p-p balance function STAR data\cite{Westfall}, 
where the solid line is the fit and dashed line is the $\rho(770)$.
The exponential is amplitude A and the dotted line is $A$ plus $A$ times 
the re-scattering of pions through the Pwave phase shift.}
\label{fig3}
\end{figure}

We turn to the central Au-Au balance function measured by STAR\cite{Westfall}.
Again using the same functions that we used above we fit the data. The solid 
line is the fit and dashed line is the $\rho(770)$. The exponential function
is amplitude A and the dotted line is $A$ plus $A$ times the re-scattering of 
pions through the Pwave phase shift. The value of $\alpha$ is 0.44 which 
means the radius of re-scattering of the pions is 0.88 fm using equation 4 with
the value of $\alpha_0$ equal to 2. We see it is a little smaller volume where 
pions can re-scatter.

\section{The dipion effective mass cocktail over a $p_t$ range in Au-Au}

In order to form this cocktail we need to consider three important ingredients 
that come into play. First is the thermal production of resonances that decay
into $\pi^+ \pi^-$ as a function of dipion $p_t$. Second is the minijet
production of $\pi^+ \pi^-$ that is not through a resonance decay and determine
its dipion effective mass spectrum as a function of dipion $p_t$. Also
we need the break up of the spectrum into partial waves. Third is rewriting 
equation 6 in a form that uses resonance or Breit-Wigner parameters (mass, 
widths) instead of phase shifts plus modify the equation to use the derived 
minijet amplitudes.

\subsection{Thermal production of resonances}

Thermal resonance production will have a Boltzmann weighting of the dipion
effective mass spectrum. Since we are projecting in $p_t$ this weighting will 
be an exponential function of the transverse mass divided by the 
temperature[6,9-13].

\begin{equation}
Weight(M_{\pi\pi})  = \frac{M_{\pi\pi}}{\sqrt{M_{\pi\pi}^2 + p^2_t}} exp \frac{-\sqrt{M_{\pi\pi}^2 + p^2_t}}{T} 
\end{equation}

This weight times the Breit-Wigner line shape is the thermal production of
the resonance which decays into the dipion system. The Breit-Wigner line
shape is given by
\begin{equation}
BW(M_{\pi\pi})  = \frac{M_{\pi\pi} M_0 \Gamma}{(M^2_0 - M^2_{\pi\pi})^2 + M^2_0\Gamma^2}.
\end{equation}
Where $\Gamma$ is the $M_{\pi\pi}$ dependent total width
\begin{equation}
\Gamma = \Gamma_0 {\frac{qB_{\ell}(q/q_s)}{M_{\pi \pi}}\over{\frac{q_0B_{\ell}(q_0/q_s)}{M_0}}}
\end{equation}
with $\Gamma_0$ being the total width at resonance, $B_{\ell}$ is the
Blatt-Weisskopf-barrier factor\cite{Hippel} for the $\ell$ of the resonance, 
$q$ is the $\pi\pi$ center mass momentum, $q_0$ is $q$ at resonance, $M_0$ is 
the mass of the resonance, and $q_s$ is center mass momentum related to the 
size(1.0 fm is used $q_s$ = .200 GeV/c).

\begin{figure}
\begin{center}
\mbox{
   \epsfysize 6.8in
   \epsfbox{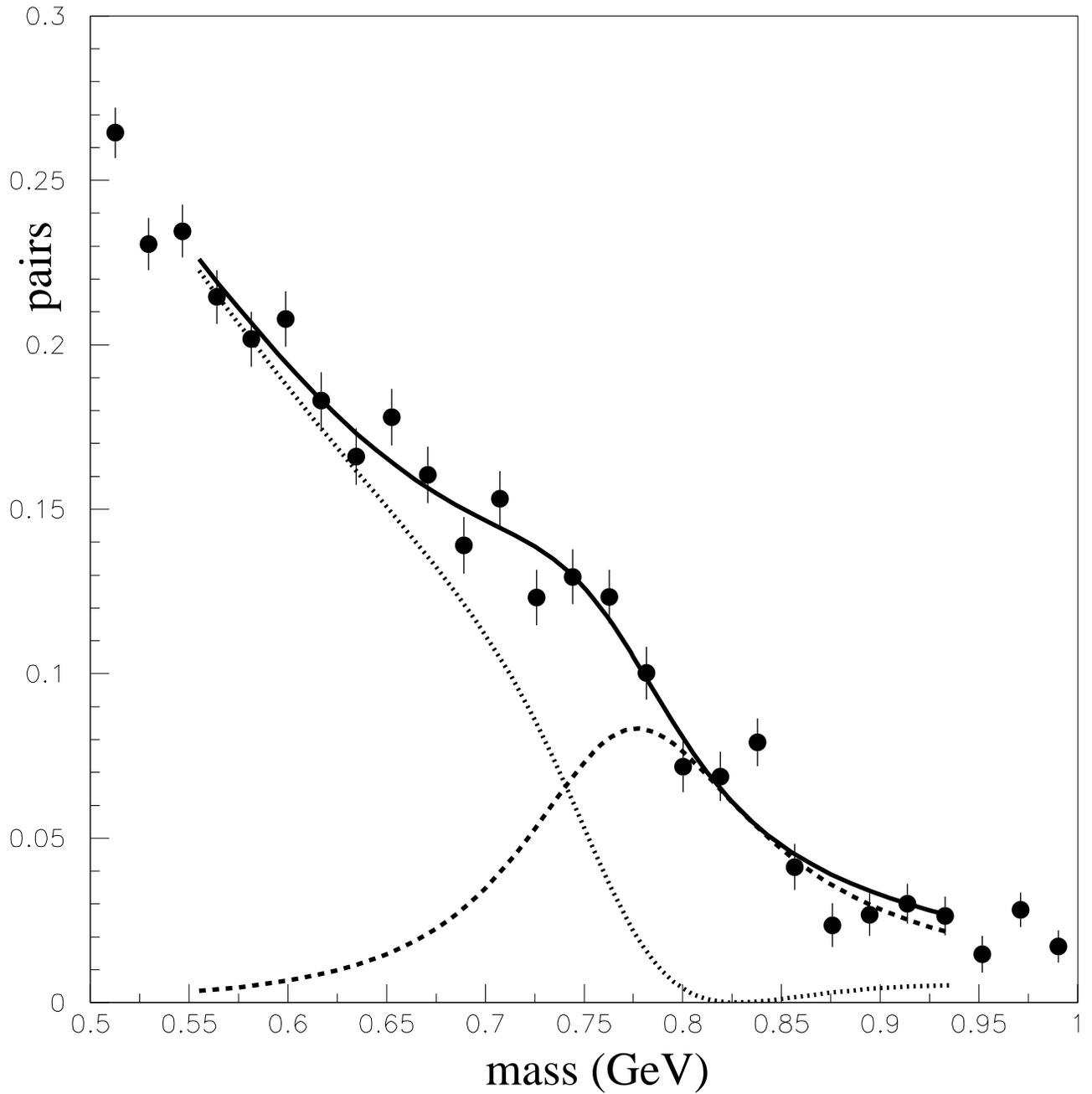}}
\end{center}
\vspace{2pt}
\caption{The fit to the Au-Au balance function STAR data\cite{Westfall}, 
where the solid line is the fit and dashed line is the $\rho(770)$.
The exponential is amplitude A and the dotted line is $A$ plus $A$ times 
the re-scattering of pions through the Pwave phase shift.}
\label{fig4}
\end{figure}

\subsection{Minijet production of dipions}

Partons that under go a hard scattering fragment into 
hadrons\cite{Trainor,pythia}. These hadrons become part of the outward flow
of hadrons with the thermal hadrons. Hadrons that have long life times will
decay outside the freeze-out volume. For these long lived resonances the dipion
spectrum will be given by the Breit-Wigner line shape of the last subsection.
Thus the source of these resonances either thermal or minijet fragmentation
will not be apparent. Resonances like the $\eta$($c\tau$ = 154000 fm), the
$\omega $($c\tau$ = 24 fm), the $\eta' $($c\tau$ = 100 fm), the 
$K^*$($c\tau$ = 4 fm), and the $\phi$($c\tau$ = 50 fm) are decaying outside
the freeze-out volume. All other resonances decay inside this volume like the
$\sigma$($c\tau$ = 1/3 fm), the $\rho(770)$($c\tau$ = 1.3 fm), and the
$f_2(1270)$($c\tau$ = 1 fm). Pions from these decays become a source for
re-scattering with pion directly produced or the ones that arise from decays.

We will use the minijet fragmentation code of PYTHIA\cite{pythia} in order to
estimate the dipion effective mass spectrum. We set up correct kinetics by
using minijets that are predicted by the program HIJING\cite{hijing}
for Au-Au collisions at $\sqrt{s_{NN}} = $ 200 GeV. We cycle through the
minijets produced by HIJING forming dipion pair($\pi^+\pi^-$) combinations 
within a given minijet fragmentation. We remove the long lived resonance listed
above because they can be handled through the direct thermal production. All 
pairs that come directly from the second class short lived resonances
are also not considered. They also can be handled through the direct thermal 
production. However combinations of their decay pions with other pions are
considered as a source of minijet dipions. We see from Figure 5 that the 
hadrons that fragment from the minijets are moving in the same direction and 
will interact with each other.  

\begin{figure}
\begin{center}
\mbox{
   \epsfysize 5.0in
   \epsfbox{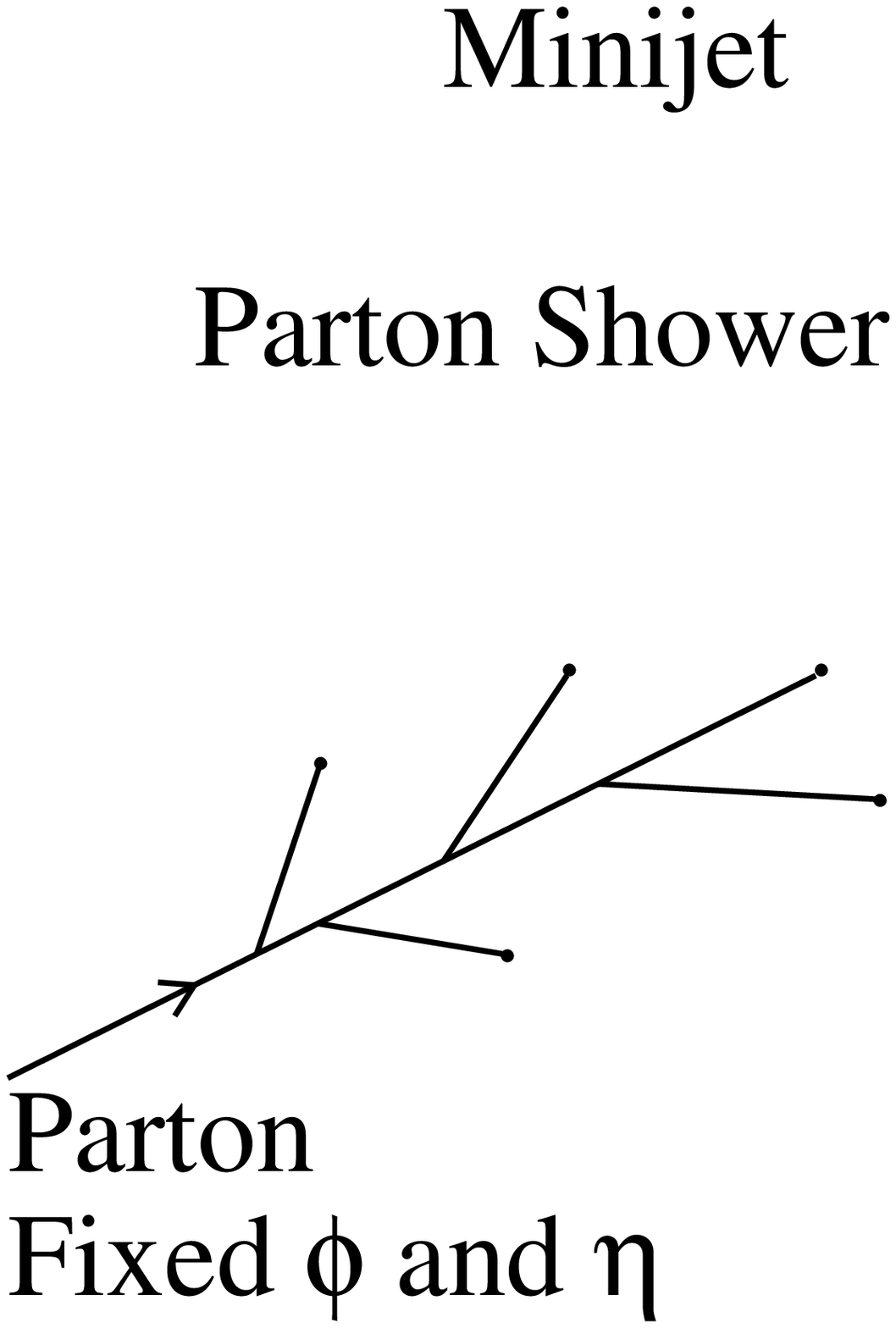}}
\end{center}
\vspace{2pt}
\caption{A minijet parton shower.}
\label{fig5}
\end{figure}

The dipions from this minijet source can be selected for its dipion $p_t$
range and decomposed into each partial wave $\ell$($\ell$wave) obtaining
an amplitude A for equation 6. We can separate out this dipion spectrum by 
using statistical and kinematic weighting. At the highest dipion mass the S, P,
D, and F waves have a $2\ell + 1$ statistical weighting. S is 1, P is 3, 
D is 5, and F is 7 making 16 units of probability. The kinematic weighting is 
given by the Blatt-Weisskopf-barrier factors. At each dipion mass 
we have a dipion spectrum from the minijets with a selection on dipion $p_t$ 
which we call $jet(M_{\pi^+\pi^-})$. Let us define Z as the ratio of the center
mass dipion momentum $q$ divided $q_s$ which is .200 GeV/c(size of 1.0 fm) all
squared 
\begin{equation}
Z = \frac{q^2}{q_s^2}.
\end{equation}

The Fwave minijet dipion weight is given by
\begin{equation}
F = \frac{7}{16}\frac{Z^3}{(Z^3+6Z^2+45Z+225)}.
\end{equation}
The Dwave minijet dipion weight is given by
\begin{equation}
D = \frac{5}{9}(1.0 - F)\frac{Z^2}{(Z^2+3Z+9)}.
\end{equation}
The Pwave minijet dipion weight is given by
\begin{equation}
P = \frac{3}{4}(1.0 - D - F)\frac{Z}{(Z+1)}.
\end{equation}
The Swave minijet dipion weight is given by
\begin{equation}
S = (1.0 - P - D - F).
\end{equation}
Thus $F(M_{\pi^+\pi^-})=Fjet(M_{\pi^+\pi^-})$, $D(M_{\pi^+\pi^-})=Djet(M_{\pi^+\pi^-})$, $P(M_{\pi^+\pi^-})=Pjet(M_{\pi^+\pi^-})$,and $S(M_{\pi^+\pi^-})=Sjet(M_{\pi^+\pi^-})$.

Let us choose a dipion $p_t$ range and plot the above minijet spectrum. 
We choose 1.6 GeV/c $<$ $p_t$ $<$ 1.8 GeV/c and plot in Figure 6 the 
Swave, Pwave, Dwave, and Fwave from minijets coming from HIJING 
for Au-Au collisions at $\sqrt{s_{NN}} = $ 200 GeV.

\begin{figure}
\begin{center}
\mbox{
   \epsfysize 6.8in
   \epsfbox{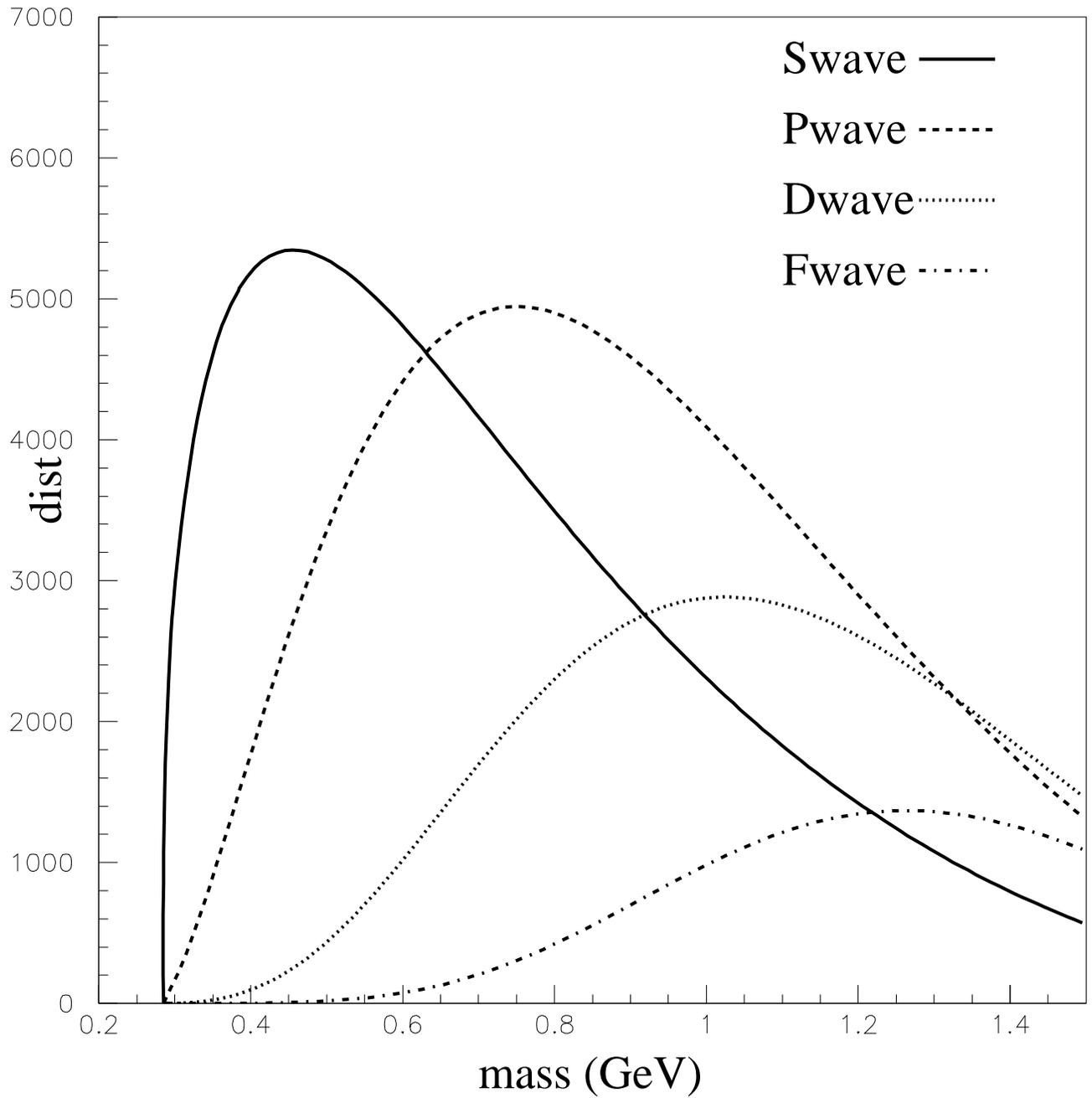}}
\end{center}
\vspace{2pt}
\caption{The dipion $p_t$ range 1.6 GeV/c $<$ $p_t$ $<$ 1.8 GeV/c showing 
the Swave, Pwave, Dwave, and Fwave from minijets coming from HIJING 
for Au-Au collisions at $\sqrt{s_{NN}} = $ 200 GeV.}
\label{fig6}
\end{figure}

\subsection{Equation 6 for Breit-Wigner parameters}

In this subsection we will alter equation 6 so it can use Breit-Wigner 
parameters (mass, width) instead of phase shifts. We will also need to
modify the re-scattering part of the equation in order to have the correct
threshold behavior we have just introduced into the minijet partial waves
above. The phase shift can be written for the $\ell^{th}$ wave as
\begin{equation}
cot\delta_\ell = \frac{(M_{\ell}^2 - M_{\pi\pi}^2)}{M_{\ell}\Gamma_{\ell}},
\end{equation}
where $M_{\ell}$ is the mass of the resonance in the $\ell$wave and 
$\Gamma_{\ell}$ is its total width.
\begin{equation}
\Gamma_{\ell} = \Gamma_{0\ell} {\frac{qB_{\ell}(q/q_s)}{M_{\pi \pi}}\over{\frac{q_{\ell}B_{\ell}(q_{\ell}/q_s)}{M_{\ell}}}}
\end{equation}
with $\Gamma_{0\ell}$ the total width at resonance, $B_{\ell}$ is the
Blatt-Weisskopf-barrier factor for the $\ell$ of the resonance, 
$q$ is the $\pi\pi$ center mass momentum, $q_{\ell}$ is $q$ at resonance, 
$M_{\ell}$ is the mass of the resonance, and $q_s$ is center mass momentum 
related to the size(1.0 fm is used $q_s$ = .200 GeV/c).

Using equation 21 we rewrite equation 6 as
\begin{equation}
|T_{\ell}|^2  = |D_{\ell}|^2 \frac{sin^2\delta_{\ell}}{PS_{\ell}} +
\frac{|A_{\ell}|^2sin^2\delta_{\ell}}{PS_{\ell}} \left| \alpha + PS_{\ell} cot\delta_\ell \right|^2
\end{equation}

The $D_{\ell}$ is the thermal production term and is constant except for the 
Boltzmann weight. The expected threshold behavior $q^{2\ell+1}$ comes from the 
$sin\delta_{\ell}$ term. Since there is  $sin^2\delta_{\ell}$ one of the
$q^{2\ell+1}$ is killed off by dividing by $PS_{\ell}$. In Figure 6 we
have put into our minijet $A_{\ell}$ the correct threshold $q^{2\ell+1}$ so
we need kill off the $q^{2\ell+1}$ of the other $sin\delta_{\ell}$ term.
Therefore equation 6 for our minijet $A_{\ell}$ we will use
\begin{equation}
|T_{\ell}|^2  = |D_{\ell}|^2 \frac{sin^2\delta_{\ell}}{PS_{\ell}} +
\frac{|A_{\ell}|^2sin^2\delta_{\ell}}{PS_{\ell}^2} \left| \alpha + PS_{\ell} cot\delta_\ell \right|^2
\end{equation}

Rewriting equation 6 for each partial wave with Breit-Wigner parameters the
first term becomes

\begin{equation}
|T_{\ell}|_1^2  = |D_{\ell}|^2 \frac{M_{\pi\pi}^2}{\sqrt{M_{\pi\pi}^2 + p^2_t}} exp \frac{-\sqrt{M_{\pi\pi}^2 + p^2_t}}{T} \frac{M_{\ell}\Gamma_{\ell}}{(M_{\ell}^2 -M_{\pi\pi}^2)^2 + M_{\ell}^2\Gamma_{\ell}^2},
\end{equation}
while the second term
\begin{equation}
|T_{\ell}|_2^2  = |A_{\ell}|^2 \frac{M_{\ell}^2\Gamma_{\ell}^2}{(M_{\ell}^2 -M_{\pi\pi}^2)^2 + M_{\ell}^2\Gamma_{\ell}^2}\left|\alpha + \frac{2qB_{\ell}(\frac{q}{q_s})(M_{\ell}^2 -M_{\pi\pi}^2)}{M_{\pi\pi}M_{\ell}\Gamma_{\ell}}\right|^2 \left(\frac{M_{\pi\pi}^2}{4q^2B_{\ell}^2(\frac{q}{q_s})}\right).
\end{equation}

\begin{equation}
|T|^2 = \sum_{\ell} |T_{\ell}|^2
\end{equation}
where
\begin{equation}
|T_{\ell}|^2 = |T_{\ell}|_1^2 + |T_{\ell}|_2^2
\end{equation}
and $|A_0|^2 = S(M_{\pi^+\pi^-})$,$|A_1|^2 = P(M_{\pi^+\pi^-})$,$|A_2|^2 = D(M_{\pi^+\pi^-})$, and $|A_3|^2 = F(M_{\pi^+\pi^-})$.

\subsection{STAR data dipion $p_t$ range (1.6 GeV/c $<$ $p_t$ $<$ 1.8 GeV/c)}

We have fitted a dipion $p_t$ range using equation 27 above for the STAR 
data Au-Au collisions at $\sqrt{s_{NN}} = $ 200 GeV 40\% to 80\% centrality.
We included minijets up to $\ell$ = 3 and resonances $\sigma$ $\ell$ = 0,
$\rho(770)$ $\ell$ = 1, and $f_2(1270)$ $\ell$ = 2. Using the arguments of 
Sec. 3 we added the $f_0$ as a direct thermal term ($|T_0|_1^2$) and only the 
$\sigma$ interfered with $\ell$ = 0 minijet background. Two other thermal
terms are present in the cocktail, the $K^0_S$ and the $\omega_0$. 

Finally the threshold effective mass region .280 GeV to .430 GeV is dominated 
by the Swave and receives contributions from minijet fragmentation, $\pi \pi$ 
Swave phase shift, $\eta$ decay, HBT adding to the like sign $\pi \pi$ 
distribution that has been subtracted away from the unlike sign $\pi \pi$ and 
the coulomb correction between the charged pions. The minijet fragmentation is 
the least known of the effects since we relied on PYTHIA, however there are 
large uncertainty in all the other effects. So for this fit we let the 
minijet fragmentation be free to fit the data and let the Breit-Wigner 
parameters for the $\sigma$ determine the Swave phase shifts plus leaving 
out all other effects. The results of this fit is shown in Figure 7. 
Table I shows the Breit-Wigner parameters used in the fit.

\bf Table I. \rm The Bret-Wigner Parameters of the fit.

\begin{center}
\begin{tabular}{|c|r|r|}\hline
\multicolumn{3}{|c|}{Table I}\\ \hline
resonance & mass(GeV) & width(GeV) \\ \hline
$\sigma$ & 1.011 & 1.015 \\ \hline
$f_0$ & 0.973 & 0.041 \\ \hline
$\rho$ & 0.748 & 0.147 \\ \hline
$f_2$ & 1.275 & 0.185 \\ \hline
\end{tabular}
\end{center}

\begin{figure}
\begin{center}
\mbox{
   \epsfysize 6.8in
   \epsfbox{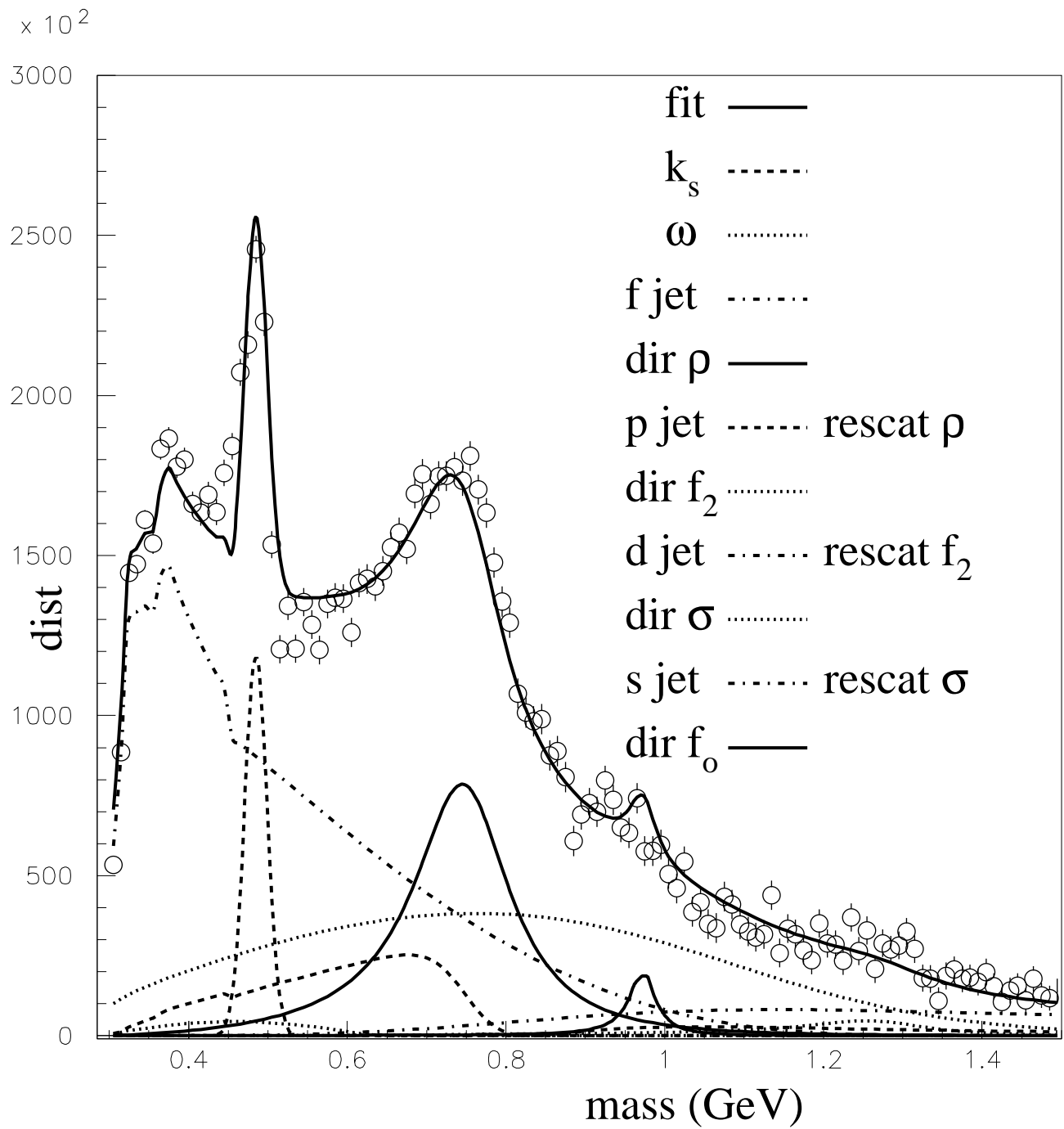}}
\end{center}
\vspace{2pt}
\caption{Fit to STAR dipion effective mass distribution (1.6 GeV/c $<$ $p_t$ 
$<$ 1.8 GeV/c) for Au-Au collisions at $\sqrt{s_{NN}} = $ 200 GeV 40\% to 80\% 
centrality using equation 27. See text for complete information.}
\label{fig7}
\end{figure}

\clearpage

The value of $\alpha$ for this $p_t$ range is .806. For coherent 
photo-production of the $\rho(770)$ from STAR data\cite{photo_rho} the value 
of $\alpha$ is 2.0 or $\alpha_0$ because the two pions emerge from a point 
source(see Appendix B). Thus the radius of the size associated with 
re-scattering is .773 fm for this $p_t$ range from equation 4. Since the 
thermal term for resonances could also come from minijet fragmentation, we 
need additional studies to determine how much production comes from minijets 
and how much comes from the fireball directly\cite{version1}.

\section{Summary and Discussion}

In this article we start with the basic definition of elastic $\pi \pi$ 
scattering. Next we show how re-scattering of pions depends on the unitary 
condition that interactions present in the phase shift of an orbital state 
must interact all the time. The process of parton fragmentation into dipion 
states through unitarity leads to a equation of production and re-scattering 
in a given orbital quantum number. This equation (equation 6) has two 
components in each orbital state: one being the thermal production of 
resonances in a dipion orbital state, the other is the re-scattering of 
dipions coming from parton or minijet fragmentation into the dipion orbital 
state which do not come directly from the resonance. Unitarity requires that 
there most be re-scatter through the resonance of the phase shift. This 
equation is used over and over again with the details presented in the 
appendix.

Equation 6 considers only elastic scattering of the $\pi \pi$ system. We 
considered the Dwave which couples to the $f_2(1270)$ ($J^{PC}$ = $2^{++}$) 
with 85\% of the cross section in the $\pi \pi$ channel. The Pwave which
coupled to the $\rho(770)$ ($J^{PC}$ = $1^{--}$) where 100\% is in the 
$\pi \pi$ channel. The Swave $\pi \pi$ ($J^{PC}$ = $0^{++}$) couples to two 
resonances the $\sigma$ and the $f_0(980)$. The $\sigma$ is purely elastic 
while the $f_0(980)$ is split between the $\pi$ $\pi$ and $K$ $\overline{K}$ 
channels. We saw that the $f_0$ was a narrow resonance. The $f_0$ resonates at 
the $K$ $\overline{K}$ threshold. Direct production of the $f_0$ gives a bump 
at the $K$ $\overline{K}$ threshold and the re-scattering of $\pi \pi$ also 
gives such the same bump at the $K$ $\overline{K}$ threshold  Sec. 3 
(Figure 2). Therefore we considered the $f_0$ as a resonance being directly
produced and decaying into $\pi \pi$ near the $K$ $\overline{K}$ threshold.

We used equation 6 to fit the dipion effective mass balance function.
We assumed the Pwave was the most important partial wave in the dipion 
effective mass range of the fit. The balance function drops off with mass so 
we assumed an exponential fall off with $M_{\pi^+ \pi^-}$. The two parts of
equation 6 where the first part being the $\rho(770)$ directly produced and 
decaying through the Pwave phase shift with the second part being the 
exponential function plus re-scattering through the $\rho(770)$. With this 
simple model we were able to fit the p-p balance (Figure 3) and the central 
Au-Au balance(Figure 4). The observed mass shifts were due to the re-scattering
effect, being $\sim$40 MeV. 

We chose a dipion $p_t$ range to do a cocktail fit to the effective dipion 
mass spectrum. We needed three important ingredients in order to do this 
fit. First is the thermal production of resonances that decay into 
$\pi^+ \pi^-$ as a function of dipion $p_t$. Second is the dipion effective 
mass spectrum as a function of dipion $p_t$ coming from minijet production 
not through resonance decay. Third we needed to rewrite equation 6 in a form 
that uses resonance or Breit-Wigner parameters (mass, widths) instead of phase 
shifts. Once these three ingredients were developed we were successful in
doing a cocktail fit (Figure 7). Since the thermal term for resonances could 
also come from minijet fragmentation, we need additional studies to determine 
how much production comes from minijets and how much comes from the fireball 
directly\cite{version1}. Other fits to more $p_t$ ranges are considered in 
Ref.\cite{version2}. 

\section{Acknowledgments}

This research was supported by the U.S. Department of Energy under Contract No.
DE-AC02-98CH10886. The author thanks William Love for the STAR analysis of the 
angular correlation data from Run 4. Also for his assistance in the production 
of figures. It is sad that he is gone.

\appendix
\section{Appendix}

Starting with equation (4.5) from Ref\cite{Aaron}
\begin{equation}
T = \frac{V_1 U'_1}{D_1} + \frac{\left( V_2 + \frac{D_{12}
V_1}{D_1} \right)\left( U'_2 + \frac{D_{12} U'_1}{D_1}\right)}{D_2
- \frac{D_{12}^2}{D_1}}
\end{equation}
In order to have the correct threshold kinematics, we define
\begin{equation}
U'_1 = U_1 \sqrt{q^{2\ell + 1}}
\end{equation}
\begin{equation}
U'_2 = U_2 \sqrt{q^{2\ell + 1}}
\end{equation}
where $q$ is the $\pi \pi$ center of mass momentum and $\ell$ is
the value of the angular momentum. The amplitude $A$ of the text
is given by
\begin{equation}
\frac{V_1 U_1}{D_1} = A
\end{equation}
Thus we have
\begin{equation}
\frac{V_1 U'_1}{D_1} = A \sqrt{q^{2\ell + 1}}
\end{equation}
The phase shift for the $\ell^{th}$ partial wave will be given by
$\delta_\ell$, where
\begin{equation}
\frac{U'_2 U'_2}{D_2} = e^{i \delta_\ell} sin\delta_\ell
\end{equation}
The above equality is true if the $D_1$ mode plays no role in the
$\pi \pi$ scattering in the $\ell^{th}$ partial wave. But in the
initial state there is a large production of $D_1$. The $U'$s are
the basic coupling of the $D'$s to the $\pi \pi$ system. In order
to decouple $D_1$ from the $\pi \pi$ system $U_1$ must go to zero.
We can maintain a finite production of $D_1$ if we define
\begin{equation}
V_1 = \frac{1}{U_1}
\end{equation}
Thus the first term in the equation becomes
\begin{equation}
\frac{V_1 U'_1}{D_1} = \frac{\frac{1}{U_1} U_1 \sqrt{q^{2\ell +
1}}}{D_1} = \frac{\sqrt{q^{2\ell + 1}}}{D_1} = A \sqrt{q^{2\ell +
1}}
\end{equation}
The form of $D_{12}$ is given by
\begin{equation}
D_{12} = \alpha U_1 U_2 + i q^{2\ell + 1} U_1 U_2
\end{equation}
$D_{12}$ is the real and imaginary part of the two pion loop from
state 1 to state 2. The $U'$s are the $\pi \pi$ couplings and the
imaginary part goes to zero at the $\pi \pi$ threshold. The $\alpha$
factor is $\alpha_0$ for re-scattering coming from a point source, but
goes to zero when re-scattering is diffractive. A simple form for $\alpha$
is given by
\begin{equation}
\alpha = (1.0 - \frac{r^2}{r_0^2}) \alpha_0
\end{equation}
where $r$ is the radius of re-scattering in fm's and $r_0$ is 1.0 fm
or the limiting range of the strong interaction.
The second term of the first equation is
\begin{equation}
\frac{\left( V_2 + \frac{D_{12} V_1}{D_1}\right) \left( U'_2 +
\frac{D_{12} U'_1}{D_1}\right)}{D_2 - \frac{D_{12}^2}{D_1}}
\end{equation}
Rewriting
\begin{equation}
\frac{\left( V_2 + \frac{\alpha U_1 U_2 V_1}{D_1} + i q^{2\ell + 1}
\frac{U_1 U_2 V_1}{D_1}\right) \left( U'_2 + \frac{D_{12}
U'_1}{D_1} \right)}{D_2 - \frac{D_{12}^2}{D_1}}
\end{equation}
Let us make substitutions
\begin{equation}
V_1 = \frac{1}{U_1}, U_2 = \frac{U'_2}{\sqrt{q^{2\ell + 1}}} ,
\frac{1}{D_1} = A, D_{12} = 0
\end{equation}
The second term becomes
\begin{equation}
\frac{\left( V_2 + \frac{A \alpha U'_2}{\sqrt{q^{2\ell + 1}}} + i
\sqrt{q^{2\ell + 1}} A U'_2 \right) U'_2}{D_2}
\end{equation}
The first term is
\begin{equation}
\frac{V_1 U'_1}{D_1} = A \sqrt{q^{2\ell + 1}}
\end{equation}
Adding the first and the second terms and substituting the phase
shift,
\begin{equation}
T = \frac{V_2}{U_2} \frac{e^{i \delta_\ell}
sin\delta_\ell}{\sqrt{q^{2\ell + 1}}} + A \left( \frac{e^{i
\delta_\ell} \alpha sin\delta_\ell}{\sqrt{q^{2\ell + 1}}} +
\sqrt{q^{2\ell + 1}} e^{i \delta_\ell} cos\delta_\ell \right)
\end{equation}
The term with the factor $\frac{V_2}{U_2}$ is the direct
production of the dipion system. We shall call this amplitude
$D$. The re-scattered amplitude is $A$ and is modified by the
dipion phase shift. These two amplitudes have some random phase
and are not coherent. Thus the cross section is
\begin{equation}
|T|^2  = |D|^2 \frac{sin^2\delta_{\ell}}{PS} +
\frac{|A|^2}{PS} \left| \alpha sin\delta_\ell + PS cos\delta_\ell \right|^2
\end{equation}

\section{Appendix}
In this appendix we determine the value $\alpha$ coming from a point 
source(thus $\alpha_0$ see equation 4 and 38) using photo-production data 
reported in Ref.\cite{photo_rho}. When a photon interacts with a strong 
field a $\rho$ or a $\pi^+$ $\pi^-$ pair is formed in a Pwave, both states 
are coherent with each other and must be added together(see equation 3 of 
Ref.\cite{photo_rho}). We write equation 3 using the phase shift 
of Pwave $\pi^+$ $\pi^-$ scattering and the width($\Gamma_0$) of the 
$\rho$ resonance. Threshold behavior is added using the phase space factor (PS)
for Pwave production (see equation 5 of Sec. 2.1).  

\begin{equation}
\frac{dN}{dM_{\pi \pi}}  = \left| A_{\rho} \frac{sin\delta e^{i \delta}}{\Gamma_0^{1/2} (\frac{PS}{PS_0})^{1/2}} + (\frac{PS}{PS_0})^{1/2} B_{\pi \pi}  \right|^2
\end{equation}

The Pwave phase factor is

\begin{equation}
PS = \frac{\frac{2q(q/q_s)^2}{(1+(q/q_s)^2)}}{M_{\pi \pi}},
\end{equation}

where $q$ is the center mass momentum of the $\pi$ $\pi$ and $q_s$ is related 
to the range of interaction of the $\pi \pi$ scattering. 1 fm is the usual
interaction distance which implies that $q_s$ is .200 GeV/c. $M_{\pi \pi}$ is 
the effective mass of the system. The Pwave phase factor at the $\rho$ 
resonance is

\begin{equation}
PS_0 = \frac{\frac{2q(q_0/q_s)^2}{(1+(q_0/q_s)^2)}}{M_{\rho}},
\end{equation}

where $q_0$ is the center mass momentum of the $\pi$ $\pi$ at the $\rho$ mass.

From equation 46 we pull out a common phase space ratio from both terms,

\begin{equation}
\frac{dN}{dM_{\pi \pi}} = \frac{1}{(\frac{PS}{PS_0})} \left| A_{\rho} \frac{sin\delta e^{i \delta}}{\Gamma_0^{1/2}} + (\frac{PS}{PS_0}) B_{\pi \pi}  \right|^2.
\end{equation}

The ratio of the $\pi$ $\pi$ production amplitude over the $\rho$ production
amplitude as averaged over world data is

\begin{equation}
\frac{B_{\pi \pi}}{A_{\rho}} = 0.85.
\end{equation}

Using equation 50 and substituting this relation for $A_{\rho}$ into equation 
49 we obtain

\begin{equation}
\frac{dN}{dM_{\pi \pi}}  = \frac{B_{\pi \pi}^2}{(\frac{PS}{PS_0})} \left| \frac{sin\delta e^{i \delta}}{0.85\Gamma_0^{1/2}} + (\frac{PS}{PS_0}) \right|^2.
\end{equation}

The second term of equation 51 can be expanded because 1 = $e^{-i\delta}e^{i\delta}$ becoming

\begin{equation}
\frac{dN}{dM_{\pi \pi}}  = \frac{B_{\pi \pi}^2}{(\frac{PS}{PS_0})} \left| \frac{sin\delta e^{i \delta}}{0.85\Gamma_0^{1/2}} + (\frac{PS}{PS_0})cos\delta e^{i \delta} -i (\frac{PS}{PS_0})sin\delta e^{i \delta} \right|^2.
\end{equation}

There is a common phase factor($e^{i\delta}$) on all three terms which has a
magnitude of 1. The magnitude for all three terms can be written as

\begin{equation}
\frac{dN}{dM_{\pi \pi}}  = (\frac{PS}{PS_0}) B_{\pi \pi}^2 sin^2\delta + \frac{B_{\pi \pi}^2}{PS(PS_0)} \left| \frac{PS_0 sin\delta}{0.85\Gamma_0^{1/2}} + PS cos\delta \right|^2.
\end{equation}

The second term has an amplitude of the form of equation 6 with

\begin{equation}
\alpha sin\delta + PS cos\delta.
\end{equation}

This implies that the maximum value of $\alpha$ is $\alpha_0$ and given by

\begin{equation}
\alpha_0 = \frac{PS_0}{0.85\Gamma_0^{1/2}} \simeq 2.0.
\end{equation}

\end{document}